\documentclass[numbers,nopreprint, superscriptaddress,floatfix]{jasatex}
\usepackage{amsmath, amssymb, amsfonts}
\begin{document}

\title[Acoustic radiation and wave-wave interactions
] {Wave-wave interactions and deep ocean acoustics
}

\author{Z. Guralnik}
\email{guralnikz@saic.com}
\author{ J. Bourdelais}
\author{X. Zabalgogeazcoa}
\affiliation{Science Applications International Corporation
1710 SAIC Drive, McLean, VA 22102%
}
\author{W.E. Farrell}
\email{wef@farrell-family.org}
\affiliation{13765 Durango Drive, Del Mar, CA 92014}
\date{\today}

\begin{abstract}
Deep ocean acoustics, in the absence of shipping and wildlife, is driven by surface processes. Best understood is the signal generated by non-linear surface wave interactions, the Longuet-Higgins mechanism, which dominates from 0.1 to 10 Hz, and may be significant for another octave.  For this source, the spectral matrix of pressure and vector velocity is derived  for points near the bottom of a deep ocean resting on an elastic half-space. In the absence of a bottom, the ratios of matrix elements are universal constants.  Bottom effects vitiate the usual ``standing wave approximation,'' but a weaker form of the approximation is shown to hold, and this is used for numerical calculations.  In the weak standing wave approximation, the ratios of matrix elements are independent of the surface wave spectrum, but depend on frequency and the propagation environment. Data from the Hawaii-2 Observatory are in excellent accord with the theory for frequencies between 0.1 and 1 Hz,  less so at higher frequencies.  Insensitivity of the spectral ratios to wind, and presumably waves, is indeed observed in the data. 
\end{abstract}

\pacs{(43.30.Nb (noise in water), 43.28.Py (Interaction of fluid motion and sound)
}

\maketitle

\section{Introduction}
Interest in deep water ambient noise has its origins in the 1930s, beginning with attempts to explain microseisms,  persistent and ubiquitous low frequency ground oscillations. Their spectrum peaks near 0.17 Hz, roughly twice the characteristic frequency of  ocean swell, and the amplitude correlates with storminess.  The most viable explanation for this phenomena, based on the work of Miche, \cite{Miche44} was originally proposed by  Longuet-Higgins \cite{LH} and extended by Hasselmann \cite{Hasselmann63} and Brekhovskikh. \cite{Brek66, Brek67} The cause is the non-linear interaction of two oppositely traveling surface  waves.  Individually, the pressure  attenuates exponentially with depth. However, wave pairs with nearly opposed wave vectors collide to excite an acoustic wave at approximately twice the frequency which propagates downward with  negligible attenuation. A review of the wave-wave mechanism and its  seismic effects,  together with citations to much of the relevant literature, can be found in Kibblewhite and Wu. \cite{K+W}

Here we give a unified exposition of the acoustic field radiated by non-linear surface wave interactions. The  power spectral density matrix (PSDM), obtained from autocorrelations and cross-correlations of pressure and vector velocity, is found both for an ocean of infinite depth, and an ocean layer over an elastic half-space. 
It is shown that the ratios of elements of the PSDM for deep observations of the acoustic signal are nearly  invariant, depending on frequency and the propagation environment but not on the details of the wave spectrum.  While the wave spectrum displays dramatic daily variability that correlates with wind speed, the propagation environment is essentially constant, aside from sound speed variations due to seasonal fluctuations or internal waves. The variability of the wave spectrum manifests itself in fluctuations in the amplitude of individual PSDM elements,  but not in their ratios.

The standing wave approximation is based on the fact that the sound speed is much greater than the phase velocity of surface gravity or capillary waves. Consequently, only pairs of surface waves with almost exactly equal frequency and almost exactly opposing propagation directions (hence almost a standing wave) can generate a non-evanescent acoustic wave, propagating without vertical attenuation (neglecting  viscosity). For acoustic waves which generate microseisms, the constraint is even more stringent; not only must the acoustic wave propagate without vertical attenuation, it must propagate nearly vertically such that the horizontal component of the acoustic wave-number is sufficiently small to excite  microseisms, which have characteristic wave-length $\sim 15$ km.

It is shown that the standing wave approximation implies that the ratios of elements of the PSDM are universal constants.  In fact, this is far too strong a result, and we do not expect its consequences to be observed, except perhaps for sufficiently lossy or rough bottoms and higher wave frequencies.  Because bottom effects negate the standing wave approximation, the radiated acoustic signal becomes sensitive to very slight deviations from the standing wave configuration.

We  show that a weaker form of the standing wave approximation will hold since the acoustic source term, unlike the acoustic propagation term, is not particularly sensitive to variations of the  horizontal wave vector $\vec k_h$, the sum of surface wave vectors, in the region for which the acoustic wave propagates without vertical attenuation, $|\vec k_h|<2\pi f/c$ ($f$ is the acoustic frequency and $c$ is the sound speed). While ratios of PSDM elements are not universal constants in the ``weak standing wave approximation,''  they will display little variability with changing surface conditions, depending only on frequency and the propagation environment but not on the details of the surface wave spectrum. 

The theory is verified with  data from the hydrophone and a pair of three-component seismometers of the \mbox{Hawaii-2} Observatory (H2O), located at 5000 m depth near 28N, 142W. \cite{Dun02}   Spectra and cross-spectra for all three-hour windows in a 50-day interval of year 2000 have been computed. Surface winds, according to the European Centre for Medium-Range Weather Forecasts (ECMWF), ranged between approximately 2 and 10 $\mathrm{m \, s^{-1}}$ during the period, and the significant wave height between 1.3 and 2.7 m. There is, indeed,  little variation in the ratios of the PSDM elements compared with the variations in the individual matrix elements; the separation between spectra, in dB, is essentially constant.  The best evidence is from coherency between pressure and vertical velocity, which is especially high for frequencies less than 1 Hz. The diminution above 1 Hz is attributed to bottom effects not embraced by the simple model.
\section{Acoustic radiation from non-linear wave interactions}
Starting with Longuet-Higgins, \cite{LH} the most common derivation of the acoustic signal radiated by wave-wave interactions involves a perturbative solution of the hydrodynamic equations. The expansion parameter characterizing the non-linearity is proportional to wave slope.  The leading term is a superposition of plane surface gravity  waves and the second-order term is an acoustic plane wave. We give an immensely abbreviated review of this approach, closely following Kibblewhite and Wu, \cite{K+W} but extending the analysis with a unified treatment both of pressure and vector velocity and simplifying it by considering the bottom as an elastic half-space.  Farrell and Munk (Ref. ~\onlinecite{Farrell+Munk3}, Appendix A) give a brief review and rationalization of the numerous solutions obtained since Longuet-Higgins.
\subsection{Perturbation equations}
The relevant perturbative solution of the irrotational hydrodynamic equations is expressed in terms of a velocity potential $\phi(\vec x,t)$ and a surface displacement $\zeta(x,y,t)$. The potential and displacement are expanded as
\begin{subequations}
\begin{eqnarray}
\phi &=& \epsilon\phi_1 + \epsilon^2 \phi_2 + \cdots \\
\zeta &=& \epsilon\zeta_{1} + \cdots
\end{eqnarray}
\end{subequations}
The expansion parameter $\epsilon$ may be set to 1 at the end of the calculation.

The first-order solution is taken to be an incompressible flow corresponding to a superposition of surface plane waves;
\begin{subequations}
\begin{eqnarray}
\lefteqn { \phi_1(\vec x,t) =} \nonumber  \\
&& \frac{1}{(2\pi)^2}\int d\vec q \frac{-i\sigma(\vec q\,)}{q}\tilde\zeta_1(\vec q\,)\exp(i\vec q\cdot\vec x_h + qz - i\sigma(q) t) \label{phi1} \\
\lefteqn{\zeta_1(x,y,t) =} \nonumber \\
&& \frac{1}{(2\pi)^2}\int d\vec q \; \tilde\zeta_1(\vec q\,)\exp(i\vec q\cdot\vec x_h - i\sigma(q) t), \label{zetafourier}
\end{eqnarray}
\end{subequations}
where $\vec x_h = (x,y),$  $\vec q = (q_x, q_y)$, $q=|\vec q\,|$ and the dispersion relation is
\begin{equation}
\sigma^2 = gq\left(1+\frac{q^2}{q_{gc}^2}\right), \; q^2_{gc}   = \frac{\rho g}{T}
\end{equation}
with surface tension $T=.074 \; \mathrm{N \;m^{-1}}$.  $q_{gc}$ is the wave number of the gravity-capillary transition. 

At next order in the $\epsilon$ expansion, the Navier-Stokes equations yield an acoustic wave equation and a surface boundary condition, with source terms dependent on the first-order solution;
\begin{subequations}
\begin{eqnarray}\label{acoustic_eqn}
&& \left(\vec\nabla^2-\frac{1}{c^2}\frac{\partial^2}{\partial t^2}\right)\phi_2 \quad = \frac{1}{c^2}\frac{\partial}{\partial t}\left((\vec\nabla\phi_1)^2\right) \\
&& \left.\left(\frac{\partial^2}{\partial t^2} \; + \; g\frac{\partial}{\partial z}\right)\phi_2\right|_{z=0} = -\left.\frac{\partial}{\partial t}\left((\vec\nabla\phi_1)^2\right)\right|_{z=0}
\label{surf_bc}
\end{eqnarray}
\end{subequations}
where $c$ is the (constant) sound speed.
\subsection{Acoustic source}
In terms of the Fourier transform of the second-order acoustic potential,
\begin{equation}
\tilde \phi_2(\omega,\vec k_h, z) =  \int dt d\vec x_h \phi_2(t, \vec x_h, z) e^{i\omega t - i\vec k_h\cdot \vec x_h} \, ,
\end{equation}
equations \eqref{acoustic_eqn} and \eqref{surf_bc}  become
\begin{subequations}
\begin{eqnarray}\label{eqso1}
&& \left ( \frac{d^2}{dz^2} + \frac{\omega^2}{c^2} - k_h^2 \right) \tilde\phi_2 = \tilde S(\omega,\vec k_h,z)\, \\ \label{eqso2}
&& \left.  \left (\frac{d}{dz} - \frac{\omega^2}{g} \right) \tilde\phi_2 \,  \right|_{z=0} =  - \frac{c^2}{g}\tilde S(\omega,\vec k_h,0)\, ,
\end{eqnarray}
\end{subequations}
where the source term is [cf. Ref. ~\onlinecite{K+W}, (4.15)]
\begin{eqnarray}
&& {\tilde{\cal S}}(\omega,\vec k_h,z) \equiv \nonumber \\
&& \int dt d \vec x_h \left (\frac{1}{c^2}\frac{\partial}{\partial t}\left((\vec\nabla\phi_1)^2\right) \right ) e^{(-i\omega t + i\vec k_h\cdot\vec x_h)} \, .
\end{eqnarray}
Using \eqref{phi1}, 
\begin{eqnarray}
&& {\tilde{\cal S}}(\omega,\vec k_h,z) = \nonumber \\
&& \frac{i\omega}{2\pi c^2} \int d\vec q\, d\vec q\,' \, \sigma(q)\sigma(q') \left (1-\frac{\vec q\cdot\vec q\,'}{qq'} \right )\tilde \zeta_1(\vec q\,)\tilde \zeta_1(\vec q\,')  \nonumber \\
&& \delta(\omega-\sigma(q)-\sigma(q'))\delta^2(\vec k_h-\vec q-\vec q\,') e^{(q+q')z} \, .
\end{eqnarray}
The $\vec q\,'$ integral is immediately evaluated because of the two-dimensional delta function. This gives 
\begin{eqnarray} \label{defS}
&& {\tilde{\cal S}}(\omega,\vec k_h,z) =  \\
&& \frac{i\omega}{2\pi c^2} \int d\vec q\, \sigma(q)\sigma(|\vec k_h-\vec q\,|) \left (1-\frac{\vec q\cdot(\vec k_h-\vec q\,) }{q|\vec k_h-\vec q\,|} \right ) \nonumber  \\
&& \tilde \zeta_1(\vec q\,)\tilde \zeta_1(\vec k_h-\vec q\,)
\delta(\omega-\sigma(q)-\sigma(|\vec k_h-\vec q\,|)) e^{(q+|\vec k_h-\vec q\,|)z} \, .\nonumber
\end{eqnarray}

\subsection{Solving for the acoustic velocity potential}
In terms of a Green's function satisfying  
\begin{subequations}
\begin{eqnarray}\label{geqn}
&& \left (\frac{d^2}{dz^2} + \frac{\omega^2}{c^2} - k_h^2 \right )G_{\omega,\vec k_h}(z,z') = -4\pi\delta(z-z')\\
&& \left . \left ( \frac{d}{dz} - \frac{\omega^2}{g} \right ) G_{\omega,\vec k_h}(z,z')\right|_{z'=0} = 0, \label{gtopbc}
\end{eqnarray}
\end{subequations}
(cf. Ref. ~\onlinecite{K+W}, Eqns. 4.30, 4.31) the solution of \eqref{eqso1} and \eqref{eqso2} is, via Green's theorem,
\begin{eqnarray}\label{theeqnfull}
\tilde \phi_2(\omega,\vec k_h,z) &=& \frac{1}{4\pi}\frac{c^2}{g}\tilde S(\omega,\vec k_h, 0)G_{\omega,\vec k_h}(z,0) \nonumber \\
&-&\frac{1}{4\pi}\int^0_{-\infty} dz' \tilde S(\omega,\vec k_h,z')G_{\omega,\vec k_h}(z,z') \, .
\end{eqnarray}
The second(bulk) term  on the right-hand side of \eqref{theeqnfull} is smaller than the first (surface) term by a factor of order $gL / c^2$,
where $L$ is a the characteristic depth scale of the surface waves, and may therefore be neglected, giving
\begin{equation}\label{theeqn}
\tilde \phi_2(\omega,\vec k_h,z) \approx \frac{1}{4\pi} \frac{c^2}{g}\tilde S(\omega,\vec k_h, 0)G_{\omega,\vec k_h}(z,0) \, .
\end{equation}
\section{The power spectral density matrix}
We introduce the 4-vector
\begin{equation}
(v_0,v_1,v_2,v_3) \equiv (\frac{P}{\rho c}, v_x, v_y, v_z),
\end{equation}
with $P$ pressure and velocity components $v_x, v_y$ in the horizontal, $v_z$ in the vertical.  The PSDM is defined by
\begin{equation}\label{PSDMdef}
M_{\mu, \nu}(\omega) \equiv \frac{1}{2\pi}\int_{-\infty}^\infty\, d\tau\, e^{i\omega\tau}\, \langle v_\mu(t)v_\nu(t+\tau) \rangle\, ,
\end{equation}
where $\langle \rangle $ denotes averaging over time.  The 10 independent elements of \eqref{PSDMdef} are evaluated by means of the far-field velocity potential $\phi_2$. With the notation
\begin{equation}
(X^0,X^1,X^2,X^3) = (ct,x,y,z) \, ,
\end{equation}
one can write
\begin{equation}\label{4veloc}
v_\mu = \frac{\partial}{\partial X^\mu}\phi_2\, .
\end{equation}
Evaluation of \eqref{PSDMdef} is thus a slight extension of the calculation which yields the spectrum of the velocity potential,
\begin{equation}\label{PhiSpectrum}
 F_{\phi_2}=\frac{1}{2\pi}\int_{-\infty}^{\infty} d\tau e^{-i\omega \tau}  \langle \phi_2(0,\vec x_h, z)\phi_2^*(\tau,\vec x_h,z) \rangle \, ,
\end{equation}
normalized such that $\int_{-\infty}^\infty d\omega F_{\phi_2}(\omega)=\langle \phi_2^2 \rangle$.

Using \eqref{theeqn} one obtains
\begin{eqnarray}\label{phi2sqr}
\lefteqn {\langle \phi_2(0,\vec x_h, z)\phi_2^*(\tau,\vec x_h,z) \rangle =\frac{1}{(2\pi)^6}\frac{1}{16\pi^2}} \nonumber \\
&& \quad \int d\omega d\vec k_h  d\omega' d\vec k'_h\,  \frac{c^4}{g^2} \langle \tilde S(\omega,\vec k_h,0)\tilde S^*(\omega',\vec k'_h,0) \rangle  \nonumber \\   
&& \quad G(\omega,\vec k_h,z)\tilde G^*(\omega',\vec k'_h,z) e^{i\omega' \tau} ,
\end{eqnarray}
where the source $\tilde S(\omega,\vec k_h,0)$ is given in \eqref{defS}.

The wave elevation spectrum is
 \begin{align} \label{defW}
F_\zeta(\vec q\,)\equiv\frac{1}{(2\pi)^2}\int d\vec x_h \langle \zeta(0)\zeta^*(\vec x) \rangle e^{i\vec q\cdot\vec x_h},
\end{align}
again normalized such that $\int d\vec q \, F_\zeta(\vec q\,)=\langle \zeta^2 \rangle $.
Equivalently, $\langle \zeta(\vec q\,)\zeta^*(\vec q\,') \rangle = (2\pi)^4 F_\zeta(\vec q\,)\delta^2(\vec q - \vec q\,')$.

Introducing the source spectrum \eqref{defS} into \eqref{phi2sqr}, using the elevation spectrum \eqref{defW}, and assuming Gaussian statistics, we get
\begin{eqnarray}
\lefteqn {\langle \phi_2(0,\vec x_h, z) \phi_2^*(\tau,\vec x_h,z) \rangle } \nonumber \\
&& =\frac{1}{16\pi^2} \int d\omega d\vec k_h \frac{\omega^2}{g^2} |\tilde G(\omega,\vec k_h,z)|^2 \nonumber \\
&& \int d\vec q\, \sigma(\vec q\,)^2\sigma(\vec k_h-\vec q\,)^2
\left(1-\frac{\vec q\cdot(\vec k_h-\vec q\,) }{q|\vec k_h-\vec q\,|}\right)^2 \nonumber \\
&&  2\, F_\zeta(\vec q\,) F_\zeta(\vec k_h-\vec q\,) \delta(\omega-\sigma(q)-\sigma(|\vec k_h-\vec q\,|)) e^{i\omega \tau} \, .
\end{eqnarray}
Note that one must be careful not to omit a combinatoric factor of 2, which arises because
\begin{eqnarray*}
\lefteqn { \langle \zeta(\vec q\,)\zeta(\vec k_h-\vec q\,)\zeta^*(\vec q\,')\zeta^*(\vec k'_h-\vec q\,') \rangle }\nonumber\\
 &&=  \langle \zeta(\vec q\,)\zeta^*(\vec q\,') \rangle \langle \zeta(\vec k_h-\vec q\,)\zeta^*(\vec k'_h-\vec q\,') \rangle  \nonumber\\
&&+ \langle \zeta(\vec q\,)\zeta^*(\vec k'_h-\vec q\,') \rangle \langle \zeta(\vec k_h-\vec q\,)\zeta^*(\vec q\,') \rangle + \cdots
\end{eqnarray*}
The remaining terms, indicated by $\cdots$, do not contribute.

We introduce the function
\begin{eqnarray}
\lefteqn {\Sigma(\omega,\vec k_h)\equiv } \nonumber\\
&& \frac{1}{8\pi^2} \frac{\omega^2}{g^2}
\int d\vec q\, \sigma(\vec q\,)^2\sigma(\vec k_h-\vec q\,)^2
\left(1-\frac{\vec q\cdot(\vec k_h-\vec q\,) }{q|\vec k_h-\vec q\,|}\right)^2\nonumber\\
&& \, F_\zeta(\vec q\,) F_\zeta(\vec k_h-\vec q\,) \delta \left (\omega-\sigma(q)-\sigma(|\vec k_h-\vec q\,|) \right ) \, ,
\end{eqnarray}
which depends on the surface wave statistics and the dispersion relation. With this definition,  the autocorrelation of the second-order velocity potential can be written
\begin{subequations}
\begin{eqnarray}
\lefteqn {\langle \phi_2(0,\vec x_h, z) \phi_2^*(\tau,\vec x_h,z) \rangle } \nonumber \\
&& = \int d\omega d\vec k_h |\tilde G(\omega,\vec k_h,z)|^2 \Sigma(\omega,\vec k_h) e^{i\omega \tau} \\
 &&=\int d\omega e^{i\omega \tau}  F_{\phi_2} \, .
\end{eqnarray}
\end{subequations}
Because of the second equivalence, which is the inverse of \eqref{PhiSpectrum}, it immediately follows that
\begin{align}
 F_{\phi_2}(\omega)=\int\, d\vec k_h
|\tilde G(\omega,\vec k_h,z)|^2 \Sigma(\omega,\vec k_h) \, .
\end{align}

Taking the appropriate derivatives, via equation \eqref{4veloc}, we get
\begin{eqnarray*}
M_{0,0} &=& \frac{\omega^2}{c^2}\int d\vec k_h\,  |G_{\omega,\vec k_h}(z,0)|^2\Sigma(\omega,\vec k_h)\nonumber\\
M_{i,j} &=&  \int  d\vec k_h\, k_{h,i} k_{h,j}  |G_{\omega,\vec k_h}(z,0)|^2 \Sigma(\omega,\vec k_h)\nonumber\\
M_{0,i} &=& -\frac{\omega}{c} \int d\vec k_h\, k_{h,i} |G_{\omega,\vec k_h}(z,0)|^2\Sigma(\omega,\vec k_h)\nonumber\\
M_{i,3} &=& i\int d\vec k_h\, k_{h,i} G_{\omega,\vec k_h}(z,0)\frac{\partial}{\partial z} G^*_{\omega,\vec k_h}(z,0)\Sigma(\omega,\vec k_h)\nonumber \\
M_{3,3} &=&  \int d\vec k_h\, \frac{\partial}{\partial z} G_{\omega,\vec k_h}(z,0)\frac{\partial}{\partial z} G^*_{\omega,\vec k_h}(z,0) \Sigma(\omega,\vec k_h)\nonumber\\
M_{0,3} &=& -i\frac{\omega}{c} \int d\vec k_h\, G_{\omega,\vec k_h}(z,0)\frac{\partial}{\partial z} G^*_{\omega,\vec k_h}(z,0)\Sigma(\omega,\vec k_h) \nonumber
\end{eqnarray*}

The following definitions lead to a more compact notation
\begin{subequations}
\begin{eqnarray}
\gimel_1 &=& |G_{\omega,\vec k_h}(z,0)|^2\Sigma(\omega,\vec k_h) \\
\gimel_2 &=& G_{\omega,\vec k_h}(z,0) \frac{\partial} {\partial z} G^*_{\omega,\vec k_h}(z,0) \Sigma(\omega,\vec k_h) \\
\gimel_3 &=&  \frac{\partial}{\partial z} G_{\omega,\vec k_h}(z,0)\frac{\partial}{\partial z} G^*_{\omega,\vec k_h}(z,0) \Sigma(\omega,\vec k_h) \\
\gimel_4 &=& \gimel_1(\vec k_h=0) \label{gimel4}
\end{eqnarray}
\end{subequations}
With these, the PSDM may be written
\begin{subequations}
\begin{eqnarray}\label{melements}
M_{0,0} &=& \frac{\omega^2}{c^2}\int d\vec k_h\,  \gimel_1 \\
M_{i,j} &=&  \int  d\vec k_h\, k_{h,i} k_{h,j} \,  \gimel_1 \quad (i,j =1,1; \, 1,2; \, 2,2)     \\
M_{0,j} &=& -\frac{\omega}{c} \int d\vec k_h\, k_{h,j} \, \gimel_1 \quad (j=1,2) \\
M_{i,3} &=& i\int d\vec k_h\, k_{h,i}  \, \gimel_2 \quad (i=1,2)  \\
M_{3,3} &=&  \int d\vec k_h\,   \gimel_3 \\
M_{0,3} &=& -i\frac{\omega}{c} \int d\vec k_h  \, \gimel_2
\label{lastmelements}
\end{eqnarray}
\end{subequations}
\section{The standing wave approximation}
The collision of two surface waves with horizontal wave-numbers $\vec q_1,\, \vec q_2$, and corresponding frequencies $\sigma(q_1),\, \sigma(q_2)$, yields an acoustic wave with frequency \mbox{$\omega = \sigma(q_1)+\sigma(q_2)$} and horizontal wave-number \mbox{$\vec k_h=\vec q_1+\vec q_2$.}  Thus, the vertical wave-number of the acoustic wave is
\begin{equation}
\gamma = \sqrt{(\omega/c)^2 - \vec k_h^2} \, .
\end{equation}
For un-attenuated propagation to the bottom,  $\gamma$ must be real, or $k\equiv|\vec k_h|<\omega / c$.
This constraint can be re-written as
\begin{equation}\label{constr}
\frac{\sigma(\vec q_1) + \sigma(\vec q_2)}{|\vec q_1 + \vec q_2|} > c \, .
\end{equation}
Given the disparity between the surface wave phase velocity and the
much larger acoustic phase velocity, $\sigma(q)/q <<c$, the constraint \eqref{constr} implies that
\mbox{$\vec q_1 \approx -\vec q_2$.} The allowed deviation from the standing wave-case,
\mbox{$\vec q_1 = -\vec q_2$,}  expressed as a variation in the relative
wavelength $ \Delta \lambda / \lambda$ or of the
propagation angle $\Delta \theta$ of one of the opposing surface waves, is
of order $c_\zeta / c$, where $c_\zeta$ is the surface wave phase velocity.
Wave spectra are essentially constant over such small variations of
angle and wave-length, which are too small to
resolve experimentally.

Thus, at sufficient depth, the integral over the horizontal wave number $\vec k_h$ in (\ref{melements} - \ref{lastmelements}) can be restricted to $k_h<\omega/c$, and the source term $\Sigma(\omega,\vec k_h)$ can be replaced with its value at $\vec k_h=0$.  In the absence of a bottom,  the squared amplitude of the Green's function $|G_{\omega,\vec k_h}(z,0)|^2$ can also be replaced with its value at $\vec k_h=0$ for the following reason.

For a bottomless ocean,  the Green's function is the solution of \eqref{geqn} with the surface boundary condition \eqref{gtopbc}. For $z<0$ ,
\begin{equation}\label{botlessgreen}
G_{\omega,\vec k_h}(z,0) = \frac{-4\pi g}{i\gamma g+\omega^2} e^{-i\gamma z}
\end{equation}
(cf. Ref. ~\onlinecite{K+W}, Section 4.2.3). In the region $k_h<\omega / c$ for which there is un-attenuated propagation to the bottom,
\begin{eqnarray}\label{squamp}
|G_{\omega,\vec k_h}(z,0)|^2 &=& \frac{16\pi^2}{\frac{\omega^4}{g^2} + \frac{\omega^2}{c^2} - k_h^2}  \nonumber \\
&=& \frac{16\pi^2 g^2}{\omega^4}\left[1 + {\cal O} (\frac{g^2}{\omega^2 c^2})\right] \, .
\end{eqnarray}

All the dependence of $|G_{\omega,k_h}|^2$ on $k_h$ is in the
sub-leading term,  which is negligible since $g /(\omega c)$ is
extremely small (about $10^{-3}$ at 1 Hz). This  is also of order $c_\zeta / c$, since for gravity waves the phase velocity is $c_\zeta=g / \sigma$ and,
for a standing wave, $\omega =2\sigma$. Note that the same small parameter, $g / \omega c$, determines the dominance of the surface term over the bulk term in \eqref{theeqnfull}. The characteristic penetration depth of
surface gravity waves is $L=1/ q = g / \sigma^2 =  4g / \omega^2$, such that the small parameter determining the dominance of the surface term over the bulk term, $gL / c^2$, is proportional to
$\left(g / \omega c \right)^2$.

The far-field spectrum  is generally computed by restricting the
integration region to $k_h<\omega / c$ and replacing both source and
propagation terms, $\Sigma_{\omega,\vec k_h}$ and $G_{\omega,\vec k_h}$, respectively,  by their values at $\vec k_h=0$. This  is known
as the standing wave approximation. In light of the above
discussion,  the standing wave approximation is a very good approximation in the absence of a bottom, or for a perfectly absorbing bottom, with errors proportional to the small parameter $c_\zeta / c$.

To make contact with well-known results, we evaluate $M_{0,0}$ in the standing wave approximation. Using  $\gimel_4$, \eqref{gimel4}, 
 \begin{eqnarray}\label{Mzz}
M_{0,0} &=& \frac{\omega^2}{c^2}\int_0^{2\pi} d\theta \int_{0}^{\omega/c} dk\, k \, \gimel_4 \nonumber\\
&=& \frac{\pi\omega^4}{c^4} \,|G_{\omega,\vec k_h}(z,0)|^2\Sigma(\omega,\vec k_h=0)\nonumber\\
&=&  \frac{\pi}{2}\frac{\omega^6}{c^4}\int d \vec q \,  F_\zeta(\vec q\,)F_\zeta(-\vec q\,) \delta(\omega-2\sigma(q))\, .
\end{eqnarray}

The wave spectrum is factored in the usual way, so $F_\zeta(\vec q\,)= F_\zeta(q)H(\theta,q)$, where $\int d\theta\, H(\theta,q)=1$, and 
\begin{equation}
I(q)\equiv\int_0^{2\pi} d\theta H(\theta,q)H(\theta+\pi,q)\, .
\end{equation}
With these definitions, equation \eqref{Mzz} becomes
\begin{equation}
M_{0,0}(\omega)=  \frac{\pi}{4}\frac{\omega^6}{c^4}\frac{q}{v}F_\zeta^2(q) I(q) \, ,
\end{equation}
where $\omega = 2\sigma(q)$, and  $v$ is the group velocity, $\partial \sigma/\partial q$. 

The pressure spectrum is then
\begin{eqnarray}\label{canonical}
F_P(\omega) &=& \rho^2 c^2 M_{0,0} \nonumber \\
&=&   \frac{\pi}{4} \left (\frac{\rho}{c}\right )^2\omega^6\frac{q}{v} F_\zeta^2(q) I(q)  .
\end{eqnarray}
This result is twice the generally adopted formula (Ref. ~\onlinecite{Farrell+Munk3}, Appendix A), indicating  a discrepancy between our approach and some previous derivations of the source term. The difference, within the spread of prior derivations, has no influence on the subsequent analysis of the standing wave approximation and the power spectral density matrix.

The implications of the standing wave approximation upon the elements of the PSDM, other than $M_{0,0}$, have not been considered previously. 

\section{Ratios of PSDM elements in the standing wave approximation}
\subsection{Perturbation theory computation}
For a perfectly absorbing bottom and surface wave pairs satisfying $k_h<\omega/c$, the wave interaction source term $\Sigma(\omega, \vec k_h)$ and the squared
amplitude of the Green's function  $|G_{\omega,\vec k_h}(z,0)|^2$ are very well approximated by their values at $\vec k_h=0$.
Thus, the PSDM elements (\ref{melements} -  \ref{lastmelements}) become
\begin{subequations}
\begin{eqnarray}\label{stwave}
M_{0,0} &=& \frac{\omega^2}{c^2} \gimel_4 \int d\vec k_h  \\
M_{i,j} &=&  \gimel_4 \int  d\vec k_h\, k_{h,i} k_{h,j} \\
M_{0,j} &=& -\frac{\omega}{c}  \gimel_4 \int d\vec k_h k_{h,j} \\
M_{i,3} &=& - \gimel_4 \int d\vec k_h\, k_{h,i} \sqrt{\frac{\omega^2}{c^2}-k_h^2} \\
M_{3,3} &=& \gimel_4 \int d\vec k_h\, \left(\frac{\omega^2}{c^2}-k_h^2\right) \\
M_{0,3} &=&  \frac{\omega}{c} \gimel_4 \int d\vec k_h \, \sqrt{\frac{\omega^2}{c^2}-k_h^2}\label{stwave_last} \,,
\end{eqnarray}
\end{subequations}
and all integrals have the limits ${k_h<\omega/c}$.

The integrals over $\vec k_h$ are trivial, giving
\begin{equation}\label{rat}
\mathbf{M} = \pi\frac{\omega^4}{c^4} \gimel_4
\begin{bmatrix}
1 & 0 & 0 & \frac{2}{3}\\
0 & \frac{1}{4} & 0 & 0 \\
0 & 0 & \frac{1}{4} & 0 \\
 \frac{2}{3} & 0 & 0 & \frac{1}{2} 
 \end{bmatrix} \, .
\end{equation}

The ratios of PSDM elements are universal constants, so we define the ratio matrix,
\begin{equation}
\label{ratios}
\mathbf{R} = \frac{M_{\mu,\nu}}{M_{0,0}} =
 \begin{bmatrix}
 1 & 0 & 0 & \frac{2}{3}\\
 0 & \frac{1}{4} & 0 & 0 \\
 0 & 0 & \frac{1}{4} & 0 \\
 \frac{2}{3} & 0 & 0 & \frac{1}{2} 
\end{bmatrix} \, .
\end{equation}
$\mathbf{R}$ is independent of both frequency and the ocean wave spectrum.   We see also that the spectrum of vertical velocity is  3 dB greater than either of the horizontals. Furthermore, the normalized pressure is exactly equal to the sum of the three velocity spectra on the diagonal. This is characteristic of any homogeneous acoustic field in the ocean. \citep{DSpain91a} 

Due to rotational invariance,  all off-diagonal components vanish except for $r_{0,3}$ and its mirror image. Sensor calibration determines the accuracy to which the elements of $\mathbf{R}$ can be estimated. A better metric is the squared coherency. This will be zero for all off-diagonal elements except $r_{0,3}$. For this term,
\begin{equation*}
\chi^2_{0,3} = r_{0,3}^2 / r_{3,3} = \frac {8}{9} \, .
\end{equation*}
The full cross-correlation matrix is
\begin{equation}\label{ChiSqMat}
\mathbf{\chi^2} = 
 \begin{bmatrix}
r_{0,0} & \frac{r_{0,1}^2}{ r_{1,1}} & \frac{r_{0,2}^2}{ r_{2,2}} & \frac{r_{0,3}^2}{ r_{3,3}}  \\
. & r_{1,1} &  \frac{r_{1,2}^2}{ r_{1,1}r_{2,2}} & \frac{r_{1,3}^2}{ r_{1,1}r_{3,3}} \\
. & . & r_{2,2} & \frac{r_{2,3}^2}{ r_{2,2}r_{3,3}} \\
. & . & . & r_{3,3}
\end{bmatrix} \, .
\end{equation}
\subsection{PSDM elements from incoherent dipoles}
The pressure field from a homogeneous surface layer of incoherent  and vertically oriented dipoles is
\begin{equation}
P(\vec x_h,z,t)=\int d\omega \int d\vec x_h^{\,\prime}  e^{-i\omega t}{ F}_\omega(\vec x_h^{\,\prime}) \partial_z\left(\frac{e^{i\frac{\omega}{c}R}}{R}\right)
\end{equation}
where ${F}$ is the source amplitude, the subscript h indicates horizontal coordinates, and
\begin{equation}
R=\sqrt{(\vec x_h-\vec x_h^{\,\prime})^2+z^2}\, .
\end{equation}
The more general case, allowing for multiple reflections at the surface and bottom, is given by Hughes. \cite{Hughes}  

The scaled pressure  spectrum, $M_{0,0}$, is
\begin{eqnarray}\label{MHu}
M_{0,0}(\omega) &=& \frac{1}{(\rho c)^2}\int d\vec x_h^{\,\prime} {\cal D}_\omega(\vec x_h^{\,\prime}) \left|\partial_z\left(\frac{e^{i\frac{\omega}{c}R}}{R}\right)\right|^2 \nonumber\\
&\approx& \frac{1}{(\rho c)^2}\int d\vec x_h^{\,\prime}  {\cal D}_\omega(\vec x_h^{\,\prime}) \left(\frac{\omega}{c}\frac{z}{R}\right)^2 \frac{1}{R^2}
 \end{eqnarray}
where ${\cal D}_\omega(\vec x_h^{\,\prime})$ is obtained from the dipole spectrum, $\langle F_\omega(\vec x)F_\omega'(\vec x^{\,\prime}) \rangle =D_\omega(\vec x\,)\delta_{\omega\omega'}\delta(\vec x-\vec x^{\,\prime})$, and we have assumed that we are at sufficient depth $z$ so that \mbox{$\omega / c >>R^{-1}$.}

Since  the dipole distribution is assumed  to be  homogenous, ${\cal D}_\omega$ is independent of $\vec x_h^{\,\prime}$.
To obtain expressions similar to those in \eqref{melements}, we make a change of variables from space to wave-number coordinates, $\vec x_h^{\,\prime}\rightarrow \vec k_h$, where
\begin{equation}
\vec k_h\equiv \frac{\omega}{c} \frac{\vec x_h^{\,\prime}-\vec x_h}{\sqrt{z^2+(\vec x_h^{\,\prime}-\vec x_h)^{2}}}
\end{equation}
or
\begin{equation}
\vec x_h^{\,\prime}-\vec x_h = z\frac{\vec k_h}{\sqrt{\frac{\omega^2}{c^2}-\vec k_h^2}} \, .
\end{equation}
Then \eqref{MHu} becomes
\begin{equation}
M_{0,0}(\omega) = \frac{1}{(\rho c)^2}{\cal D}_\omega  \int_{|\vec k_h|<\frac{\omega}{c}} d\vec k_h \, .
\end{equation}
Using $P=\rho\dot\phi$ and  $v_i={\partial \phi} / {\partial x^i}$, we obtain the rest of the PSDM elements:
\begin{subequations}
\begin{eqnarray}\label{Humelements}
M_{0,0}(\omega) &=& \frac{1}{(\rho c)^2}{\cal D}_\omega \int d\vec k_h  \\
M_{i,j}(\omega) &=& \frac{c^2}{\omega^2} \frac{1}{(\rho c)^2}{\cal D}_\omega \int d\vec k_h\, k_{h,i} k_{h,j} \\
M_{0,i}(\omega)&=& - \frac{c}{\omega} \frac{1}{(\rho c)^2}{\cal D}_\omega \int d\vec k_h k_{h,i}  \\
M_{i,3}(\omega) &=& - \frac{c^2}{\omega^2} \frac{1}{(\rho c)^2}{\cal D}_\omega \int d\vec k_h\, k_{h,i} \sqrt{\frac{\omega^2}{c^2}-k_h^2}  \\
M_{3,3}(\omega) &=& \frac{c^2}{\omega^2} \frac{1}{(\rho c)^2}{\cal D}_\omega \int d\vec k_h\, \left(\frac{\omega^2}{c^2}-k_h^2\right)  \\
M_{0,3}(\omega) &=&  \frac{c}{\omega} \frac{1}{(\rho c)^2}{\cal D}_\omega \int d\vec k_h \, \sqrt{\frac{\omega^2}{c^2}-k_h^2}\, ,
\end{eqnarray}
\end{subequations}
and each integral has the limits ${k_h<\omega/c} $

This agrees precisely with the result \eqref{stwave}--\eqref{stwave_last} obtained using perturbative non-linear wave interaction theory in the standing wave approximation, provided that the surface dipole spectrum is related to the wave spectrum by
\begin{equation}
{\cal D}_\omega=\rho^2\omega^2 \gimel_4 \, .
\end{equation}

\section{Bottom effects}
In general, the effect of the bottom is a modification of the Green's function appearing in (\ref{melements}-\ref{lastmelements}).  The standing wave approximation which leads to \eqref{rat}   is dependent on the squared amplitude of the Green's function of the operator
\begin{equation}
{\hat {\cal L}}=\frac{d^2}{dz^2} + \frac{\omega^2}{c^2} - k_h^2
\end{equation}
being a nearly constant function of $k_h$ in the region \mbox{$k_h<\omega / c$.}  The fact that this holds in the bottomless case is a lucky accident. In general, the Green's function is non-analytic at $k_h=\omega / c$, which is the transition point from vertical propagation to attenuation, so there is no reason to expect its squared amplitude to be constant in a neighborhood of this point.  Upon inclusion of a bottom,  the squared amplitude of the Green's function  becomes a rapidly varying function on the interval $k_h=[0,\omega / c]$. The location of the peaks of this function are determined by solutions of the eigenvalue equation
\begin{equation}
\left ( \frac{d^2}{dz^2} + \frac{\omega^2}{c^2}-k_{h,l}^2 \right ) \phi=0\, ,
\end{equation}
subject to the appropriate top and bottom boundary conditions. On the real interval $k_h=[0,\omega / c]$, the Green's function has narrow peaks centered about real $k_{h,l}$ corresponding to normal modes and wider peaks centered about the real part of complex $k_{h,l}$ corresponding to leaky modes (see, for example, Ref. ~\onlinecite{Kats12}, Section 2.1).
\subsection{Green's function for a layer over a half-space}
We wish to obtain the Green's function $G_{\omega,\vec k_h}(z,z_s)$ satisfying
\begin{equation}
\left(\partial_z^2 + \frac{\omega^2}{c^2}-k_h^2\right)G_{\omega,\vec k_h}(z,z_s)=\delta(z-z_s)\, ,
\end{equation}
subject to the surface boundary condition (for the case of gravity waves),
\begin{equation}
\left.\left(\partial_z-\frac{\omega^2}{g}\right)G\right|_{z=0}=0
\end{equation}
and the mixed Dirichlet-Neuman (impedance) bottom boundary condition,
\begin{equation}\label{dirneum}
\left.\left(G+h(\vec k_h)\partial_z G\right)\right|_{z=z_{\rm bottom}} =0\, .
\end{equation}
The function $h(\vec k_h)$ depends on the characteristics of the region below $z_{\rm bottom} \equiv d$.

Assuming constant sound speed $c$ in the water column above $z_{\rm bottom}$, and taking $z_s=0$ (the surface), the solution is 
\begin{eqnarray}\label{grsoln}
\lefteqn {G_{\omega,\vec k_h}(z,0) = } \nonumber \\
&& \frac{ g\left(e^{-i\gamma(d+z)} + R(\omega, \vec k_h) e^{i\gamma(d+z)}\right)}
  {(\omega^2+i\gamma g)e^{-i\gamma d} + R(\omega, \vec k_h) (\omega^2-i\gamma g)e^{i\gamma d}}\, ,
 \end{eqnarray}
where
\begin{equation}
\gamma\equiv \sqrt{\frac{\omega^2}{c^2} - \vec k_h^2}\, ,
\end{equation}
and $R$ is the reflection coefficient;
\begin{equation}
R=\frac{M-1}{M+1}
\end{equation}
with $M=Z_b/Z$, and the top and bottom layer impedances are defined by
\begin{subequations}
\begin{eqnarray}
Z &=& \frac{\rho\omega}{\gamma}\\
Z_b &=& i \rho\omega  h(\vec k_h)\label{botimped}\, .
\end{eqnarray}
\end{subequations}

The function $h(\vec k_h)$ entering the bottom boundary condition \eqref{dirneum} is obtained from the matching condition across the top--bottom interface or across multiple interfaces if the bottom is layered. For a fluid half-space bottom,  one matches the vertical velocity $v_z$ on either side of the interface, while for
a solid half-space bottom, one matches $v_z$ and the components of the stress tensor containing a vertical index, $\tau_{iz}$. This has been done before in numerous places (see for example Ref. ~\onlinecite{Brek60}).  Without derivation,  the result for a fluid layer over an elastic half-space is given below.

For the fluid half-space with density $\rho_b$ and sound speed $c_b$,
\begin{equation}\label{fluiddirneum}
h(\vec k_h) = -i\frac{\rho_b}{\rho \gamma_b},\qquad \gamma_b \equiv \sqrt{\frac{\omega^2}{c_b^2}-\vec k_h^2}\, .
\end{equation}

For the elastic half-space with density $\rho_b$, shear wave velocity $c_s$, and compression wave velocity $c_p$,
\begin{eqnarray}\label{soliddirneum}
h(\vec k_h) &=& i\frac{c_s^2}{\omega^4 \rho}\left[\frac{\vec k_h^2-\gamma_s^2}{\gamma_p}\left( (2\mu+\lambda)\gamma_p^2+\lambda \vec k_h^2\right)
-4\mu \vec k_h^2\gamma_s\right]\nonumber \\
\gamma_p &\equiv& \sqrt{\frac{\omega^2}{c_p^2}-\vec k_h^2},\qquad
\gamma_s \equiv \sqrt{\frac{\omega^2}{c_s^2}-\vec k_h^2}
\end{eqnarray}
where $\mu$ and $\lambda$ are the Lam\'{e} parameters, related to the shear and compression wave velocities by
\begin{eqnarray}
c_p &=& \sqrt{\frac{\lambda + 2\mu}{\rho_b}}\nonumber\\
c_s &=& \sqrt{\frac{\mu}{\rho_b}}\, .
\end{eqnarray}

\subsection{Numerical example}
\begin{figure}[h]
\noindent\includegraphics[width=20 pc]{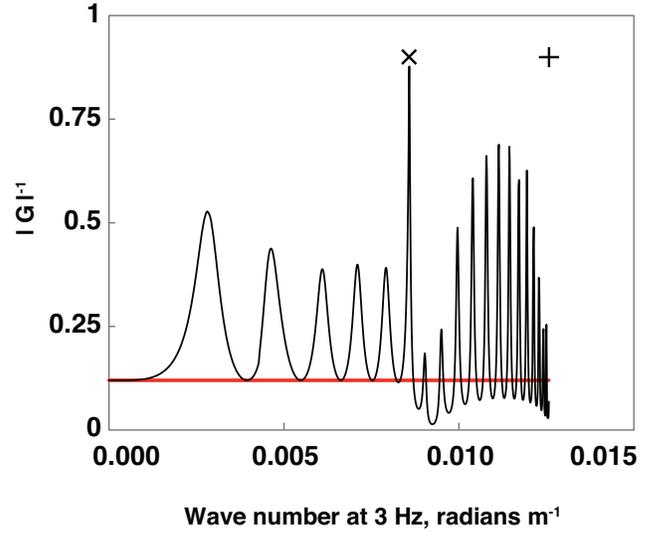}
\caption{The squared amplitude of the Green's function $G_{\omega,k_h}(0,z_{\rm bot})$ on the interval $k_h=[0,\omega / c]$ for a fluid layer over an elastic half-space at 3 Hz. Red shows the result for a bottomless ocean, which  is very nearly flat on this interval. The cutoff wavenumber, $\omega / 1500 = .0126$, is denoted by $+$; the critical wavenumber, $k_{h, {\rm crit}} = \omega / 2200 = .0086$, with $\times$.
}
\label{GSQ}
\end{figure}

As a specific example,  Green's function for the model given in Table \ref{T1} was calculated for an acoustic frequency of 3 Hz (Fig. \ref{GSQ}). A power loss of 6 dB per bounce, independent of angle, is assumed for $k_h>k_{h, {\rm crit}}$ \mbox{($k_{h, {\rm crit}} = \omega / c_s = .0086$)}. 
\begin{table}[htdp]
\caption{Model used for calculating power spectrum matrix}
\begin{center}
\begin{tabular}{ c c c }
\hline
\hline
  & Ocean layer & Half-space  \\
\hline
$\rho \; \mathrm{(kg \, m^{-3})}$&1000 & 2000 \\

$c_p \; \mathrm{(m \, s^{-1})}$& 1500 & 4400  \\

$c_s \; \mathrm{(m \, s^{-1})}$&  & 2200  \\

thickness (m) & 5000 & $\infty$ \\
 \hline
 \hline
\end{tabular}
\end{center}
\label{T1}
\end{table}
%

The figure shows pronounced narrow peaks, corresponding to normal modes on the interval \mbox{$k_h=[k_{h,{\rm crit}},\omega / c]$} where the reflection coefficient has modulus nearly equal to 1 and acoustic energy is not lost by propagation through the bottom layer.  The critical wave number is $\omega/c_s$ where $c_s$ is the shear wave velocity in the half-space. A small power loss per bottom bounce is incorporated to give the normal modes $k_{h,l}$  a small imaginary component such that these peaks have finite height.

There are also a series of wider peaks on the interval $k_h=[0,k_{h,{\rm crit}}]$, corresponding to leaky modes. A  general feature of bottom effects is that the acoustic signal increases over the bottomless case.  Moreover, the PSDM ratios can be expected to differ from \eqref{ratios} due to oscillations of the squared amplitude of the Green's function which violate the standing wave approximation; it is no longer possible to make the replacement $G_{\omega,\vec k_h} \rightarrow G_{\omega,\vec k_h=0}$ inside the integrals in \eqref{melements}.

\section{The Weak Standing Wave Approximation}
Although the squared amplitude of the Green's function is not flat on the interval $k_h=[0,\omega / c]$ when bottom effects are considered, the surface wave spectrum remains unaltered by bottom effects, except possibly for the organ-pipe modes. \cite{LH, Hasselmann63}
With this limitation, one can  make the replacement
 \begin{equation}
 \Sigma(\omega,\vec k_h)\rightarrow \Sigma(\omega,\vec k_h=0),
 \end{equation}
 in (\ref{melements}-\ref{lastmelements}), which we call the
``weak standing wave'' approximation.  Since the wave spectrum, $\Sigma$, is evaluated at $\vec k_h=0$, it can be taken outside those $\vec k_h$ integrals, giving
\begin{equation}
M_{\mu,\nu} = \Sigma(\omega,\vec k_h=0) \Xi_{\mu,\nu}
\end{equation}
where
\begin{subequations}
\begin{eqnarray}
\Xi_{0,0} &=& \frac{\omega^2}{c^2}\int d\vec k_h\,  |G_{\omega,\vec k_h}(z,0)|^2 \label{chi00} \\
\Xi_{i,j} &=&  \int  d\vec k_h\, k_{h,i} k_{h,j}  |G_{\omega,\vec k_h}(z,0)|^2  \\
\Xi_{0,j} &=& -\frac{\omega}{c} \int d\vec k_h k_{h,j} |G_{\omega,\vec k_h}(z,0)|^2 \\
\Xi_{i,3} &=& i\int d\vec k_h\, k_{h,i} G_{\omega,\vec k_h}(z,0)\frac{\partial}{\partial z} G^*_{\omega,\vec k_h}(z,0)  \\
\Xi_{3,3} &=&  \int d\vec k_h\, \frac{\partial}{\partial z} G_{\omega,\vec k_h}(z,0)\frac{\partial}{\partial z} G^*_{\omega,\vec k_h}(z,0)  \\
\Xi_{0,3} &=& -i\frac{\omega}{c} \int d\vec k_h\, G_{\omega,\vec k_h}(z,0)\frac{\partial}{\partial z} G^*_{\omega,\vec k_h}(z,0) \, ,
\end{eqnarray}
\end{subequations}
and each integral has the limits $k_h<\omega/c$.

The  term that depends on the wave spectrum, $\Sigma$, cancels in the ratios of PSDM elements, which therefore depend solely on $\Xi_{\mu,\nu}$. In the absence of signal other than that generated by surface wave interactions, similar arguments imply that
the dependence on the wave spectrum also cancels in other ratios of deep ocean acoustic correlation functions.

Thus we propose that the PSDM element ratios in deep water for the signal  generated by wave interactions, while not universal constants, are independent of the details of the wave spectrum. They will depend solely on frequency and the propagation environment, which together determine the form of the Green's function $G_{\omega,\vec k_h}$.
\subsection{Pressure near the bottom}
The effect of the bottom on acoustic pressure is expressed as the ratio of the $M_{0,0}(z)$ matrix elements. Using the weak standing wave approximation,  \eqref{chi00}, the relative effect of the bottom on pressure is given by
\begin{equation}\label{PressureRatio}
\frac{M^{\,b}_{0,0}(z)}{M_{0,0}(z)} = \frac{\Xi_{0,0}^{\,b} (z)}{\Xi_{0,0}(z) } 
\end{equation}
where $()^b$ indicates there is a bottom under the model ocean. 

\begin{figure}[h]
\includegraphics[width=20 pc]{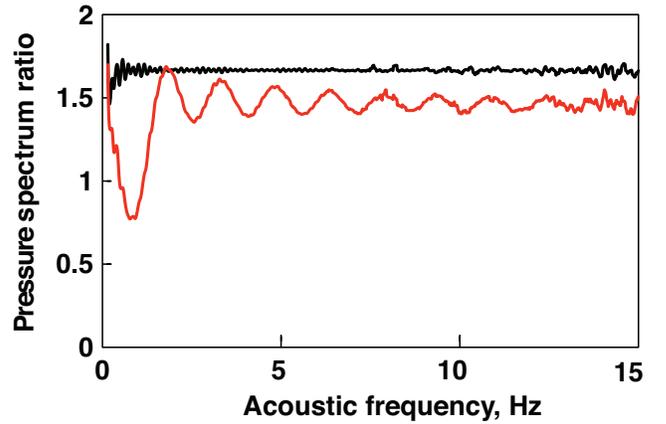}
\caption{The effect of the solid half-space under the ocean is a uniform increase  in bottom pressure by the factor 1.66 (black), or 2.3 dB. The pressure increase is slightly less, 1.46,  500 m above the bottom (red).}
\label{B}
\end{figure}

Evaluating \eqref{PressureRatio} for two depths, we find the solid bottom raises the pressure at the bottom by an average of 2.3 dB, and at 500 m off the bottom the increase is 1.6 dB (Fig. \ref{B}). These values are averages over frequencies less than 15 Hz, but the figure shows the smoothed spectrum is flat. Kibblewhite and Wu, using a more detailed elastic model, found the bottom lifted the pressure spectrum by 3 dB  (Ref. ~\onlinecite{K+W},  p. 116 and Fig. 7.8). The spectra plotted in the figure are equivalent to the function $B$ introduced by Farrell and Munk  to express the effect of the bottom on the spectrum of acoustic pressure arising from wave-wave interactions on the ocean surface [Ref. ~\onlinecite{Farrell+Munk3}, (4)].
\subsection{Model spectra on the ocean floor}
\begin{figure}[h]
\includegraphics[width=20 pc]{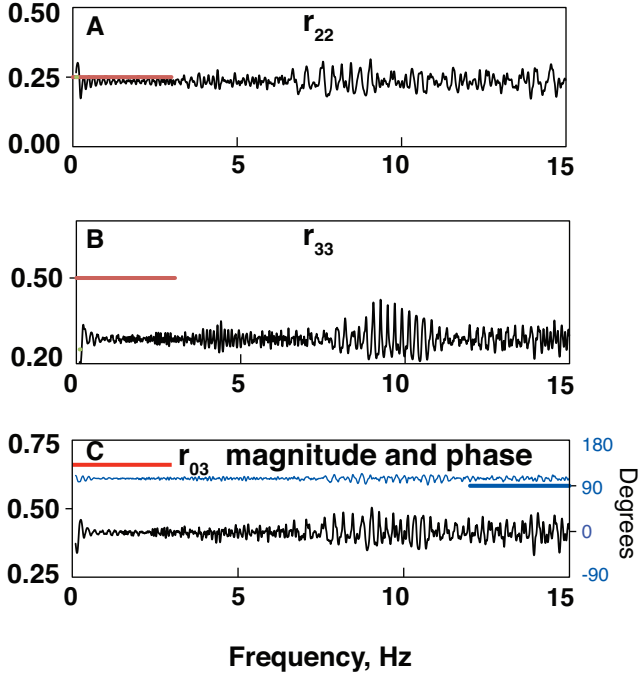}
\caption{PSDM  element ratios $r_{\mu,\nu}=\Xi_{\mu,\nu} / \Xi_{0,0}$  (black) for a point just above the bottom for a 5,000 m water layer over an elastic half-space (Table \ref{T1}).  The short red (blue) line segments are the ratios for a bottomless ocean.
The red  segments terminate at 3 Hz, the reference frequency of the Green's function shown in Fig. \ref{GSQ}.
}
\label{Rbot}
\end{figure}

We have calculated all elements of  $\Xi_{\mu\nu}$ for the simple half-space model  (see Table 1).  For an observation point at the bottom but in the water, the matrix elements shown in Fig. \ref{Rbot} are obtained.  Smoothing the elements  over plotted band, 0-15 Hz,  gives the ratio matrix,
\begin{equation}\label{Rhat}
 \mathbf{R(z_{bot})} =
\begin{bmatrix}
 1 & 0 & 0 & -0.1 + 0.4i \\
 . &  0.23 & 0 & 0   \\
 . & .  & 0.23 & 0  \\
 .   & .  & . & 0.29
\end{bmatrix} \, .
\end{equation}
When smoothed over twice the bandwidth the results are essentially the same. If the bottom loss is halved, to 3 dB, the changes in the matrix are small. Element $r_{3,3}$ is reduced by 10\%, the other two diagonal elements are raised by 10\%, and the magnitude of $r_{0,3}$ becomes 25\% less.

The principal effect of the bottom is to reduce  the vertical velocity, relative to pressure, by about 40\% with respect to the case of the bottomless ocean. There are two consequences: both  $r_{3,3}$ and $r_{0,3}$ are smaller by about this amount. In addition,  the coherency between pressure and vertical velocity is less:  $\chi^2_{0,3}=.58$, vs. 0.89 for the bottomless ocean. The diagonal sum of $\mathbf{R}$ is 1.75, the difference from 2 reflecting the inhomogeneity of the acoustic field near the bottom. \cite{DSpain91a}

\subsection{Model spectra near the bottom}
\begin{figure}[ht]
\includegraphics[width=20 pc]{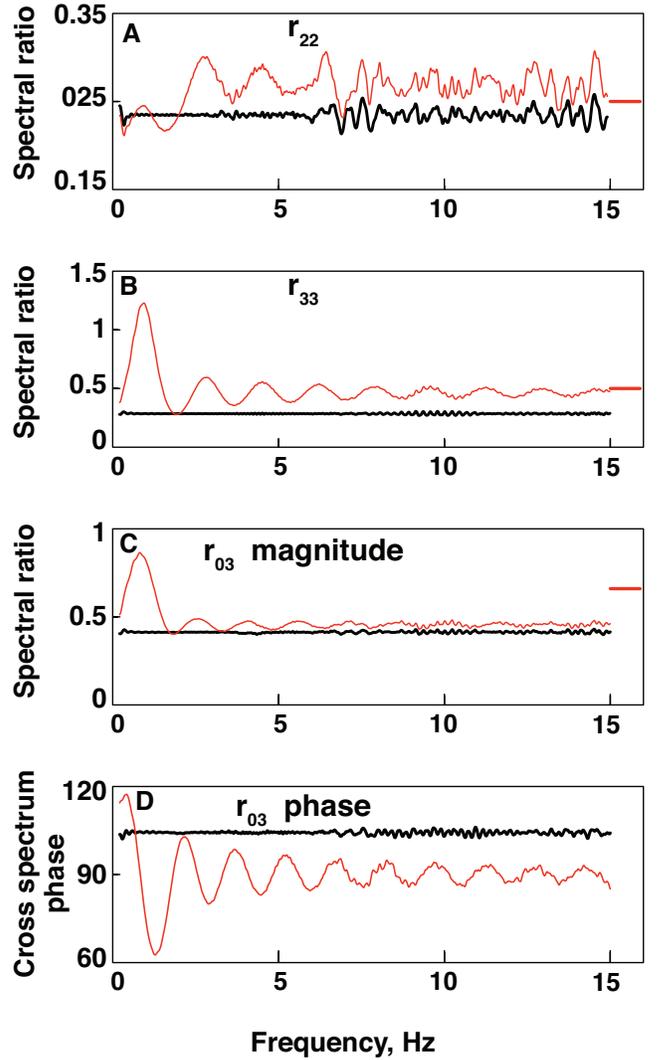}
\caption{PSDM element ratios (smoothed over 0.3 Hz) at the ocean bottom (black, recapitulating Fig. \ref{Rbot}) and 500 meters above the bottom (red). 
For the diagonal elements and frequencies towards the right, the red spectra  approach the bottomless ocean result, denoted by the red line segments on the right margin (panels A, B). The off-diagonal element does not (panels C, D).
}
\label{Rdeep}
\end{figure}

As the observation point rises to 500 m above the bottom, 10\% of the water depth, the spectra evolve as shown in Fig. \ref{Rdeep}. Smoothing the spectra as before gives the following ratio matrix:
\begin{equation}\label{RhatB}
 \mathbf{R(0.9 \, z_{bot})} =
\begin{bmatrix}
 1 & 0 & 0 & 0.0+.48i\\
 . &  0.27 & 0 & 0   \\
 . & .  & 0.27 & 0  \\
 .   & .  & . & 0.49
\end{bmatrix} \, .
\end{equation}
The vertical velocity element, $r_{3,3}$, has nearly reverted to the bottomless value (36), but, surprisingly, the $r_{0,3}$ element is 25\% less.
The diagonal sum is 2.03, reflecting the near  homogeneity of the acoustic field. 

\section{Evidence in data from station H2O}
The theory is applied to bottom data from station H2O, \citep{Dun02} located at 5,000 meters in  a thin-sediment area of the Pacific, mid-way between Hawaii and California.  The instrumentation consists of a buried (0.5 m) Guralp seismometer (Guralp System Limited's CMG-3) and Geospace geophone (GTC, Inc.'s Geospace Technologies\textsuperscript{\texttrademark} HS-1), and a hydrophone a little off the bottom.  The two velocity sensors differ in the way the motion of the inertial mass is sensed: the Guralp is better at very low frequencies, the Geospace at very high, but they are comparable over our analysis band. In addition, assimilated surface winds for the H2O location were provided by the European Centre for Medium-range Weather Forecasts from the ERA-interim (ECMWF Re-analysis) results.

H2O data (hydrophone, channel HDH; Geospace, channels EHZ, EH1, EH2;  Guralp, channels HHZ, HL1, HL2), obtained from the Incorporated Research Institutions for Seismology (IRIS) Data Management Center, have been studied for years 2000-2002, inclusive. Spectra were calculated for three-hour windows to a resolution of 0.1 Hz, giving about 2000 equivalent degrees of freedom. The spectra have been examined for the whole interval, but this discussion is restricted to days 200-250 in year 2000. Spectra were calibrated according to the nominal transfer functions on file at the IRIS Data Management Center. Various small inaccuracies were discovered and corrected. For example, there were discrepancies at the times the gain was changed. The spectra for the vertical component of the Geospace geophone were raised by 3 dB. This makes them consistent with spectra from the vertical component of the Guralp seismometer,  is compatible with the instrument noise model, and corrects  a demonstrable error in its nominal generator constant.

\subsection{Bottom acoustics and surface winds}
It is generally observed that bottom acoustics, from frequencies less than 1 Hz to frequencies above 30, are highly correlated with surface wind. \cite{Farrell+Munk2}  Typical are the profiles of the acoustic power at 0.5 Hz (Fig. 5)  and 1.76 Hz (Fig. 6). The lower frequency was chosen because the coherency between pressure and vertical velocity is greatest. The higher was selected to be on a relative extremum of the coherency and away from any sediment resonance.
\begin{figure}[ht]
\includegraphics[width=20 pc]{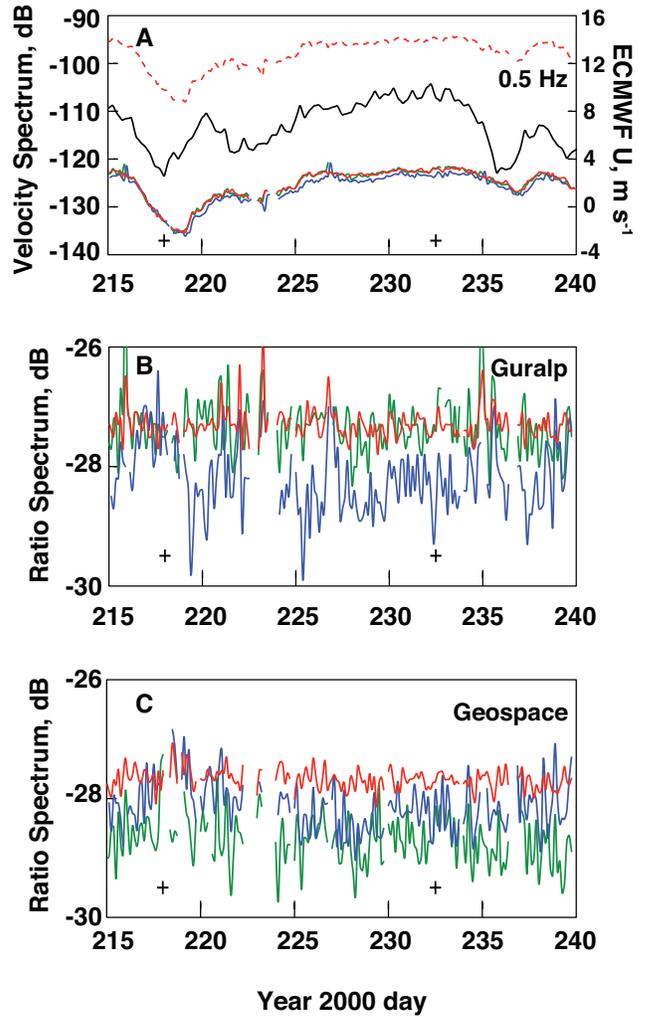}
\caption{Profile of acoustic power at 0.5 Hz  for Guralp data with dash red pressure (scaled by $(\rho c)^2$), and red, blue, green denoting the vertical and two horizontal components of velocity (A).  ECMWF wind (black) is referenced to the right axis. Panels B and C are the the three ratios of velocity to scaled pressure for the two instruments.  Blanks indicate sections where there were no data or where the value was dropped on a first difference screen. The plus symbols at the bottom of each panel denote windows at days  218.125 ($U=3.3$) and 232.125 ($U=10.1$) for which detailed results are presented below. 
}
\label{0.5Profile}
\end{figure}
\begin{figure}[ht]
  \noindent\includegraphics[width=20 pc,angle=0]{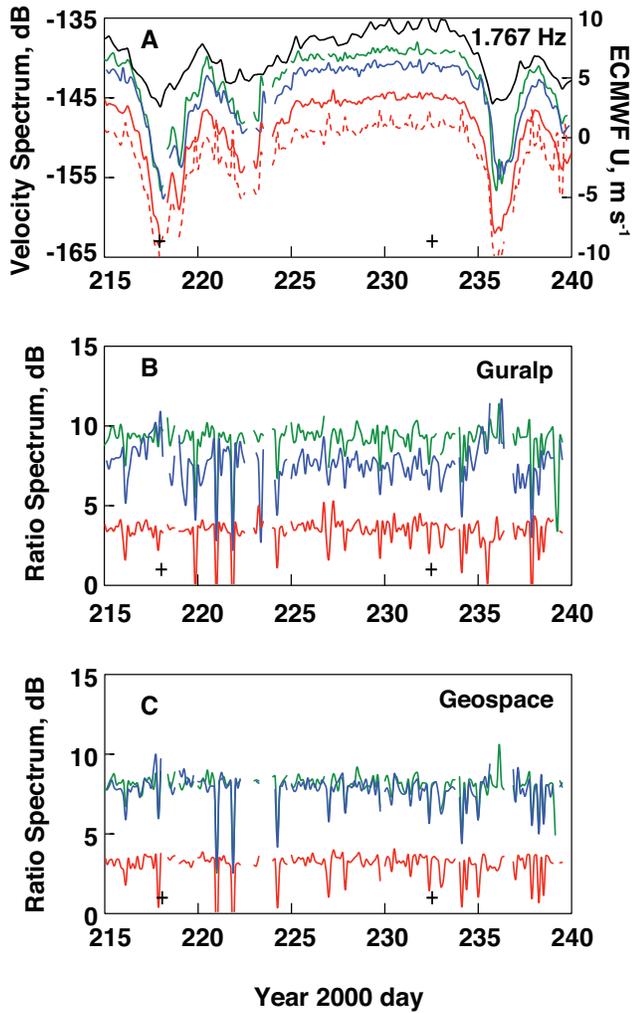}\\
  \caption{Profiles of acoustic power at 1.76 Hz  for Guralp data (colored) and ECMWF wind (black) (A). Panels B and C are the the three ratios of velocity to scaled pressure for the two instruments as in Fig. 5. See its caption for additional details. Note that the acoustics in Panel A more closely follow the dips in the wind than is the case at the lower frequency of Fig. \ref{0.5Profile}. 
   }
  \label{1.7Profile}
\end{figure}

The vertical scales in Fig. \ref{0.5Profile}, Panel A, have been adjusted so that the variation in wind (black) is about as large as the variation in acoustics (colored). In Panel A of Fig \ref{1.7Profile}, the wind has about half the range of the acoustics. Comparing axes, the slope in log acoustic power is roughly $\mathrm{2.5 \, dB / (m \, s^{-1})}$ at these frequencies.  It has previously been shown (Ref. ~\onlinecite{Farrell+Munk3}, Fig 2) that between 6 and 30 Hz the slope of the spectrum is in the range 2.7 - 2.9 $\mathrm{dB /(m \, s^{-1})}$.

In both figures, the flatness of the components of the velocity ratios  (Panels B, C) shows that scaling vector velocity by normalized pressure is effective at reducing or even eliminating the correlation  with surface wind. This is as expected from the foregoing theory, to the extent the wind is a proxy for the waves.

\subsection{Off-diagonal matrix elements}
The off-diagonal elements of the PSDM are the most diagnostic of the wave-wave acoustic field because, expressed as squared coherency (see Eq.  \ref{ChiSqMat}), they are independent of sensor calibration. The element $\chi^2_{0,3}$ is the only off-diagonal element that does not vanish for the elementary models we are testing. 

\begin{figure}[ht]
\includegraphics[width=20 pc]{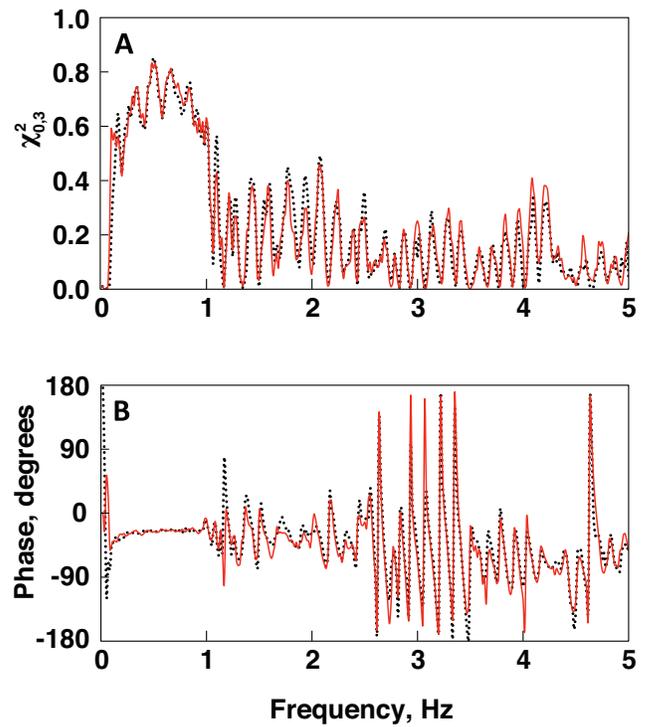}
\caption{Pressure and vertical velocity are highly correlated for frequencies less than 1 Hz (A), with phase angle of $-25^\circ$ at 0.5 Hz  (B, Table \ref{T2}). The coherency at higher frequencies, though less, is still significant, but because of the lower coherency, the phase fluctuates more. These Guralp exemplars include a high-wind window (red, day 232.125, U=10.1 $\mathrm{m \, s^{-1}}$) and a low-wind window (black dash, day 218.125, U=3.3 $\mathrm{m \, s^{-1}}$). 
}
\label{GuralpX03}
\end{figure}

The spectrum of $\chi^2_{0,3}$ (Fig. \ref{GuralpX03}) shows a profound discontinuity at 1.03 Hz, which is just below the gravest sediment resonance (see Fig. 10). The two effects are presumably related.  For lower frequencies the measured coherency is even higher than the model (0.8 vs 0.58, see Eqs. 37 and 62). The phase of the cross-spectrum (Fig \ref{GuralpX03}, Panel B) in the high-coherency band is about $-25^\circ$, far from the model expectation of $104^\circ$ (62). However, the phase of the cross-spectrum depends on the phase response of both instruments, and, as explained in the Discussion, there are questions about the hydrophone transfer function.

At higher frequencies the coherency wobbles around 0.1-0.2, and there are large swings in phase. There is pronounced rippling in the coherency, with a periodicity of approximately 6 cycles/Hz. These are likely organ pipe modes, a possibility previously considered but not embraced. \citep{Stephen07}  

\begin{figure}[ht]
\includegraphics[width=20 pc]{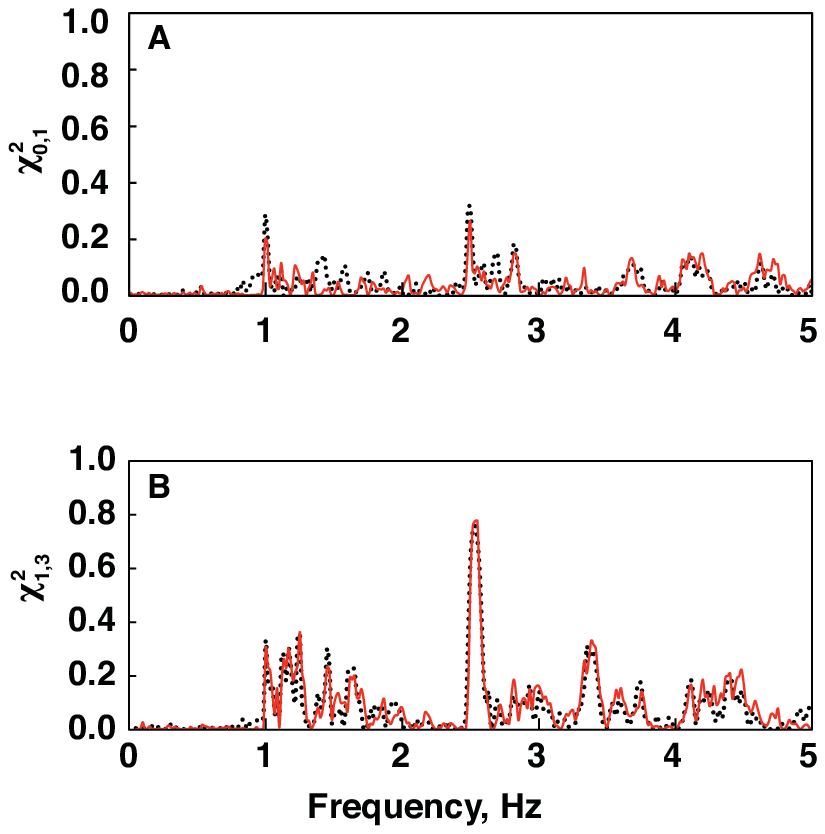}
\caption{For the same windows displayed in Fig. \ref{GuralpX03}, the coherency between pressure and horizontal velocity (A) and between vertical and horizontal velocity (B) is low everywhere, but especially for frequencies less than 1 Hz,  where $\chi^2_{0,3}$ is highest (see Table \ref{T2}). 
}
\label{GuralpX13}
\end{figure}

In theory, the other five off-diagonal elements vanish; in practice, they nearly do so for $f < 1$ Hz and are not large at higher frequencies (Fig. \ref{GuralpX13}). The spike in $\chi^2_{1,3} $ near 2.5 Hz corresponds to one of the P-SV modes in the sediments (see Fig. \ref{Diagonals}).

\begin{figure}[ht]
\includegraphics[width=20 pc]{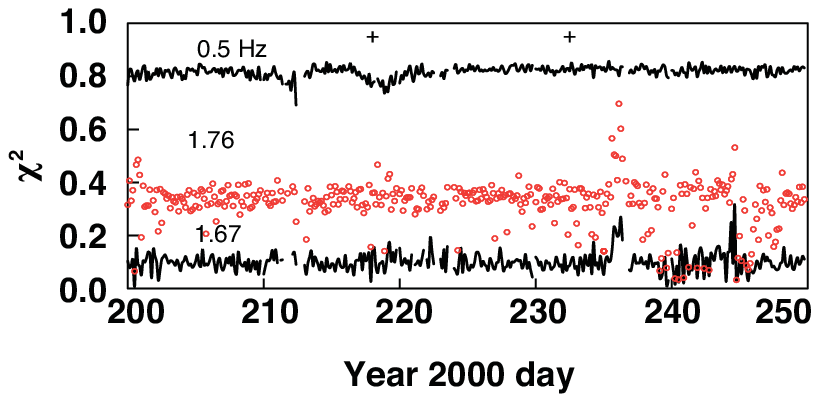}
\caption{The coherency between pressure and vertical velocity (Geospace sensor, three representative frequencies) is virtually constant, irrespective of wind speed.  For 0.5 Hz, there is a slight dimple near day 218 when the wind was low. Results for 1.67 Hz around day 235 are unusual, as is the scatter between days 240 and 245 at 1.76 Hz.  (Gaps are due to missing data or spectra dropped on a first difference criterion.)
}
\label{GeospaceX03Histories}
\end{figure}

Dividing the velocity spectra by the (scaled) pressure  (Figs 5 and 6, panels B and C) is effective at obliterating  the wind correlation. Thus, it is not surprising that the cross-spectrum of pressure and vertical velocity is equally quiescent (Fig. 9). Another perspective on the residual wind signature is visible in comparing the pairs of spectra plotted in Figs. 7 and 8. Red is the coherency under a strong wind, and black dash under a weak wind. The pairs of curves are virtually indistinguishable.
\subsection{Diagonal matrix elements}
\begin{figure}[ht]
\includegraphics[width=20 pc]{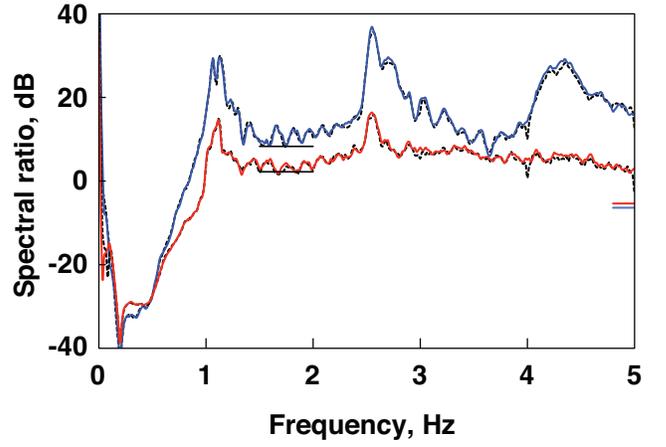}
\caption{Guralp matrix elements $r_{1,1}$ (blue) and  $r_{3,3}$ (red) for day 232.125 ($U= 10.1  \, \mathrm{m \, s^{-1}}$). The adjacent black dash curves are the spectra for day 218.125 ($U= 3.3 \,\mathrm{m \, s^{-1}}$). The black bars are averages of the power at 1.76 Hz, exemplified in Panel B of Fig. \ref{1.7Profile}. The short bars on the right are matrix values from \eqref{Rhat}. Spectra for the Geospace instrument are similar.
}
\label{Diagonals}
\end{figure}

The power spectra of the velocity components down the diagonal of the PSDM (Fig. \ref{Diagonals}) are relatively flat for frequencies above 1 Hz but fall precipitously below that. Excluding sediment resonances, the spectra are in accord with the spot measurements obtained from smoothing the results at 0.5 Hz and 1.76 Hz displayed in Figs. \ref{0.5Profile} and \ref{1.7Profile}, respectively, and listed in Table \ref{T2}. The steep drop below 1 Hz we tentatively attribute to a hydrophone calibration error (see Discussion). The bumps on the spectra of both components at 1 Hz and 2.5 Hz  are attributed to P-SV sediment resonances, as is the broader peak at 4.25 Hz  for $r_{1,1}$ (blue). \citep{Stephen07}
\subsection{Summary}

The model yields a PSDM ratio matrix for which all  elements are independent of both frequency and  wave spectrum. This is true whether they are expressed in natural units or as squared coherency.

The measured PSDM ratio matrix is, indeed, insensitive to wind, which we take as a proxy for waves, but strongly depends on frequency.  There is a sharp transition at 1 Hz.  For lower frequencies, the coherency of the off-diagonal elements is closer to the model than for higher. A sharp transition in the diagonal matrix elements occurs at this  same critical frequency. Below it, the ratio elements plunge by 30 dB, which is entirely caused but the sharp rise in the pressure spectrum.  Although we have suggested this is a calibration matter, the correspondence in frequency of the two effects suggests they are linked in the physics. Above the transition frequency, the diagonal elements are flatter, which is more in accord with the model. The variations are within 5 dB. Peaks of up to 10 dB, which are attributed to P-SV resonances in the sediments, are discounted. However, the levels are much larger than the model results, an effect also attributed to energy trapped in the sediments.

The important numeral results are highlighted in Table \ref{T2}. For six representative matrix elements (column 1), this shows the model values (columns 2) and the observed values (columns 3-6) for both sensors and the two representative frequencies.

\begin{table}[htdp]
\caption{Selected elements of the spectral ratio matrix at 0.5 and 1.76 Hz for both H2O seismic sensors.  The 12 values for the diagonal elements $r_{i, \,j}$ were calculated by averaging the spectrum over the 50 days, for which shorter segments are plotted in Figs. 5 and 6. The values of the off-diagonal elements were calculated by similar smoothing in time at the appropriate frequencies (for the Guralp, see Figs. 7 and 8). 
}
\begin{center}
\begin{tabular}{ c c | c  c  | c c c }
\hline
\hline
&  Theory & \multicolumn{2}{ c  }{0.5 Hz} &\multicolumn{2}{ c }{1.76 Hz} \\

&  \eqref{Rhat}  & Guralp & Geospace & Guralp & Geospace \\
\hline

$r_{1,1}$ & 0.23 &   $1.86\times10^{-3}$ & $1.39\times10^{-3}$ & 8.29 & 6.28 \\
$r_{2,2}$ & 0.23 &   $1.54\times10^{-3}$ & $1.57\times10^{-3}$ & 5.82 & 6.12\\
$r_{3,3}$ & 0.29 &   $1.86\times10^{-3}$ & $1.70\times10^{-3}$ & 2.21 & 2.07 \\
 \hline
 $r_{0,1}$ &  0 & $<.01$  &  $<.01$ & $<.01$ & $<.01$ \\
 $r_{0,3}$ &    $-0.1+0.4i $& $ .81 - .43i$ &$ .82 - .30i$ & $.32 -.48i$ & $.32 - .85i$  \\
  $r_{1,3}$ &   0 &  $<.01$ & $<.01$ & $<.02$  & $<.02$  \\
 \hline
 \hline
\end{tabular}
\end{center}
\label{T2}
\end{table}

For the diagonal elements (top rows), the values for 0.5 Hz are orders of magnitude smaller than the model, and the values at 1.76 Hz (and above) are 10 to 40 times larger.  The off-diagonal elements (bottom three rows) fit the model better at 0.5 Hz than at 1.76, although in neither case has the result been smoothed over the prominent ripples. 

The frequency dependence observed in the coherency of the off-diagonal matrix elements is  inconsistent with the universal constants predicted by the standing wave approximation in the case of an ocean layer resting on an elastic half-space.  Thus, other effects, possibly bottom scattering, are influencing the wave-generated sound at this site.  On the other hand, the insensitivity to wind, as predicted by the weak standing wave approximation, is strongly upheld by these data.
\section{Discussion}
We have shown that  the theoretical predictions of the standing wave approximation are extremely strong, yielding PSDM ratios which are universal constants.  While it is possible that these PSDM ratios are observed under certain conditions,  such as a high loss bottom, they are certainly not observed in the H2O data.  

Bottom effects vitiate the standing wave approximation, leaving in place a weaker version of the approximation, the predictions of which are observed at H2O.  In particular, the PSDM ratios depend on frequency, but are insensitive to changes  in the surface conditions.  Direct comparison of the PSDM ratios to an elastic half-space model show less agreement.  This is due, presumably, to the oversimplification of the bottom and propagation conditions in the model, as well  as to instrument calibration error.
\subsection{Bottom interaction}
The primitive model of the ocean's bottom replicates the essential features of more elaborate approaches. Pressure is increased a few dB because of the bottom, and this increase is uniform over the band 1-30 Hz, at least. The augmentation diminishes slightly as the observation depth lessens.  Given a quantitative estimate for the influence of the bottom, overhead wave properties can be inferred from (corrected) deep pressure observations through application of (33), although there is a factor 2 discrepancy between this result and some other theories. The model of the ocean bottom will need to be extended to account for the layer of ocean sediments  before velocity observations can be used for the same purpose.

Pressure and vertical velocity are most coherent for frequencies below 1 Hz; indeed, the observed coherence exceeds the model result. The coherence falls above 1 Hz.  The sediments overlying the basement were not modeled, but the decrease in coherence is  attributed to P-SV energy trapped in this layer.  This same effect can explain the weak but significant rise at 1 Hz  in the coherence between horizontal velocity and both pressure and vertical velocity. 

We view the peaks in the velocity spectra as evidence of vertically polarized shear waves, naturally excited when a pressure wave in an acoustic medium is incident on an elastic medium. The extremely low shear velocity of the surficial sediments has two effects: the transmitted SV rays are nearly perpendicular to the boundary and they are strongly polarized in the horizontal direction. The resonance occurs when the rays are efficiently reflected by internally layering in the sediments. Zeldenrust and Stephen \cite{Zeldenrust} applied  the theory of Godin and Chapman \cite{Godin99} to interpret the resonances at H2O as evidence of a chert layer approximately 13 m below the bottom, about in the middle of the 30 m of sediments.  Scattering is a further complication.   Any zones in the elastic medium with strong impedance contrasts and rough boundaries will scatter both SV and SH energy.

\subsection{Influence of the ocean's sound speed profile}
Taking the ocean to have constant sound speed ignores the refraction due to the sound channel (e.g. Ref. ~\onlinecite{Farrell+Munk3}, Fig. 3). However, for a surface layer of incoherent dipoles, the bottom signal is dominated by the source region with diameter six times the water depth (Ref ~\onlinecite{Farrell+Munk2}, Eq. 7). 
\subsection{Inference of the wind-wave spectrum}
Bottom acoustic observations are beginning to be used to estimate the spectrum of ocean surface waves. To do this, within the framework of acoustic radiation from wave-wave interactions,  two corrections are necessary. Allowance must be made for the bottom interaction and the overlap integral. 

The correction of the pressure spectrum for bottom interaction appears to be straightforward. The bottom elevates the pressure a few dB at the bottom, with a slight decrease moving up the water column away from the bottom.  The bottom effect, in fact, is smaller than the factor-of-two discrepancy between our derivation and some previous results.

It is just as significant that the bottom correction does not depend on frequency.  Thus, the theory for the bottomless ocean can be used to infer the slope of the wave spectrum from the slope of the pressure spectrum \cite{Farrell+Munk3} with no further corrections. 

Spectra of bottom velocity at this site are contaminated by sediment effects for $f \gtrapprox 1$ Hz. Horizontal is more affected than vertical, and models incorporating the sediments will be required to back out this contamination. Data at lower frequencies may still be applicable. However, all bottom sensors are equally useful for relating changes in the spectrum to changes in overhead waves.
\subsection{Sensor calibration}
\subsubsection{Hydrophone}
When the nominal transfer function of the H2O HDH hydrophone is adopted, spectral ratios of velocity to (scaled) pressure, drop precipitously for frequencies less than 1 Hz (Fig. \ref{Diagonals}). At 0.5 Hz, the observed spectral ratios for both seismic sensors are more than 100 times smaller than theory (Table \ref{T2}). 

There is more evidence that the nominal gain of the hydrophone is too high at low frequencies. Velocity spectra at low frequencies have been calculated from ECMWF directional wave models, and they are in reasonable agreement with the H2O seismic observations (Ref. ~\onlinecite{Farrell+Munk3}, Fig. 4). Pressure spectra calculated from the same models are orders of magnitude less than observations.

In addition, the (scaled) pressure spectrum at 0.5 Hz varies between -95 dB and -105 dB at H2O, depending on overhead wind (e.g. Figs. ~\ref{0.5Profile}, ~\ref{1.7Profile}). Observations at the Aloha Cabled Observatory, scaled similarly, range between -113 and -128 dB, some 20 dB less. \cite{Duennebier12}
\subsubsection{Seismometers}
The similarity between the spectra of all six seismic channels at low frequency is in accord with the model (cf. panels AA in Figs. \ref{0.5Profile}, \ref{1.7Profile}, and Fig. \ref{Diagonals}) and indicates  relative calibrations accurate to a dB (once spectra of the Geospace vertical have been lifted 3 dB). However, the phase of the cross-spectrum between pressure and vertical velocity differs for the two instruments for frequencies less than 1 Hz. Sediment effects preclude applying these checks at higher frequencies (e.g. panels BB in Figs. ~\ref{0.5Profile}, \ref{1.7Profile}).

\begin{acknowledgments}
Walter Munk has been a constant inspiration, and we have greatly benefited from innumerable conversations with Chuck Spofford and Brian Sperry. We again acknowledge the extraordinary accomplishment of F. Duennebier and the entire H2O team for installation and operation of the H2O system and thank the IRIS Data Management Center for data curation. We are grateful to  Jean Bidlot at the European Center for Medium-Range Weather Forecasts for several custom data sets extracted from the ECMWF database. This work was partially supported by the Office of Naval Research.
\end{acknowledgments}

\bibliographystyle{jasanum}

\end{document}